\documentclass[conference,a4paper]{APSIPA2026}
\usepackage{amsmath}
\usepackage{graphicx}
\usepackage{multirow}
\usepackage{threeparttable}
\usepackage{siunitx}
\usepackage{bm}
\usepackage{cleveref}
\usepackage{makecell}
\usepackage{subfig}
\usepackage{amsfonts}
\usepackage{multirow}
\usepackage{booktabs}
\usepackage{makecell}
\usepackage[
  backend=biber,
  bibstyle=ieee,
  citestyle=numeric-comp,
  sorting=none,
  sortcites=true]{biblatex}

\usepackage{geometry}
\usepackage{fancyhdr}

\fancypagestyle{firststyle}{
  \fancyhf{}
  \fancyhead[C]{2026 Asia Pacific Signal and Information Processing Association Annual Summit and Conference (APSIPA ASC)}
}

\newcommand{\R}{\mathbb{R}}
\newcommand{\C}{\mathbb{C}}
\newcommand{\Fo}{F_{\text{o}}}

\crefname{figure}{Fig.}{Figs.}
\Crefname{figure}{Figure}{Figures}

\begin{document}

\title{Lead Vocal Separation from Vocal Ensemble Mixtures Using Phoneme Alignment}

\author{
\authorblockN{
Yuma Narahata\authorrefmark{1}\authorrefmark{2}, 
Tomohiko Nakamura\authorrefmark{2}, 
Yuki Saito\authorrefmark{1}, 
Hiroshi Saruwatari\authorrefmark{1}
}

\authorblockA{
\authorrefmark{1}
The University of Tokyo, Tokyo, Japan}

\authorblockA{
\authorrefmark{2}
National Institute of Advanced Industrial Science and Technology (AIST), Tokyo, Japan}
}

\maketitle
\thispagestyle{firststyle}
\pagestyle{empty}

\begin{abstract}
Contemporary a cappella singing often has a lead-and-accompaniment texture, where the lead vocal (Vo) part carries the main melody and the remaining vocal parts provide accompaniment.
Owing to their distinct roles, separating the Vo part from the remaining vocal parts, referred to as Vo separation, enables downstream applications such as lyric recognition and minus-one accompaniment generation for vocal ensemble music.
Despite these potential applications, acoustic cues for this task are limited because the target and interfering sources are all singing voices with similar acoustic characteristics and often overlap in time, making Vo separation challenging.
In this paper, we propose a Vo separation model that uses phoneme alignment of the Vo part as auxiliary information.
The proposed model is based on band-split RoPE Transformer (BS-RoFormer), a state-of-the-art music source separation model, and introduces frame-level phoneme labels into its intermediate representations using feature-wise linear modulation (FiLM).
Experimental results show that phoneme-alignment conditioning improves Vo separation performance over an audio-only baseline and yields larger average gains than conditioning only on Vo singing/silence activity.
Further analysis suggests that the advantage of phoneme-label information is larger when fewer remaining vocal parts share the same phoneme as Vo.
\end{abstract}

\section{Introduction}
Contemporary a cappella singing is a form of group singing performed using only human voices and body sounds~\cite{acappella101}.
Compared with traditional forms of group singing, such as choral singing, it often adapts popular-music recordings into vocal ensemble performances in which each vocal part is assigned to a single singer.
It is also characterized by lead-and-accompaniment textures (i.e., a single melody line with accompaniment lines) and instrumental imitation~\cite{Duchan2007Collegiate}.

In contemporary a cappella singing, the lead vocal (Vo) part carries the main melody and lyrics, whereas the remaining vocal parts provide accompaniment and percussive sounds.
Owing to this difference in musical roles, separating these components would enable different applications for the two outputs.
The separated Vo part could provide a cleaner input for lead-voice processing in vocal ensemble recordings, such as lyric recognition and singing voice conversion.
The remaining parts, in turn, could serve as Vo-minus-one accompaniments, extending karaoke-style applications from instrument-accompanied popular music to vocal ensemble music.
Thus, we address a two-stem task that separates a vocal ensemble mixture into the Vo part and the remaining vocal parts: \emph{lead vocal separation}, or \emph{Vo separation}.

Vo separation is closely related to vocal ensemble separation, which estimates individual vocal parts from a mixture recording.
While earlier studies on vocal ensemble separation have focused on choral and barbershop recordings~\cite{Petermann2020ISMIR,Sarkar2021Interspeech}, recent studies have targeted contemporary a cappella recordings~\cite{TNakamura202306ICASSP,Pan2025AI4Music,Luca2026ICASSP}.
These studies provide an audio-only basis for separating vocal ensemble recordings.
Unlike previous vocal ensemble separation studies that estimate individual vocal parts, Vo separation defines the target by its musical role rather than by source type because both the Vo part and the remaining parts consist of human voices.
Thus, an audio-only model must infer the Vo/non-Vo distinction using only acoustic cues in the mixture.
This motivates the use of role-specific auxiliary information that can specify the Vo part more directly.

To explore auxiliary information for Vo separation, we draw inspiration from lyric-informed singing voice separation~\cite{MeseguerBrocal2020ISMIR,Gupta2022IEEEACMTASLP}.
In singing voice separation, the target singing voice carries the main melody and lyrics, whereas the accompaniment provides the remaining musical components.
The Vo part plays an analogous role in vocal ensemble mixtures.
This analogy suggests that lyric information can serve as a role-specific cue for specifying the Vo part within a vocal ensemble mixture.
However, unlike instrument-accompanied music mixtures, vocal ensemble mixtures may include intervals where the remaining vocal parts share the same lyrics or phonemes with the Vo part.
It is thus not obvious how effectively lyric-derived information can distinguish the Vo part from the remaining vocal parts.
This motivates us not only to use phoneme alignment as auxiliary information but also to analyze how its effect depends on the phonetic overlap between Vo and the other parts.

In this paper, we propose a Vo separation method that uses time-aligned phoneme information (phoneme alignment) as auxiliary information.
The proposed model introduces feature-wise linear modulation (FiLM)~\cite{Perez2018AAAI} into band-split RoPE Transformer (BS-RoFormer)~\cite{Lu2024ICASSP}, a state-of-the-art music source separation model, to condition intermediate representations on frame-level phoneme labels of the Vo part.

We further investigate how phoneme alignment contributes to Vo separation in six-part vocal ensemble recordings from a six-part vocal ensemble corpus.
To separate the effect of phoneme labels from Vo activity, we compare the proposed method with an activity-only condition using singing/silence labels.
We also analyze separation performance under different phoneme-overlap conditions between Vo and the other parts to examine when phoneme alignment remains effective.

\section{Related Work}
\begin{figure}[t]
  \centering
     \includegraphics[width=\linewidth]{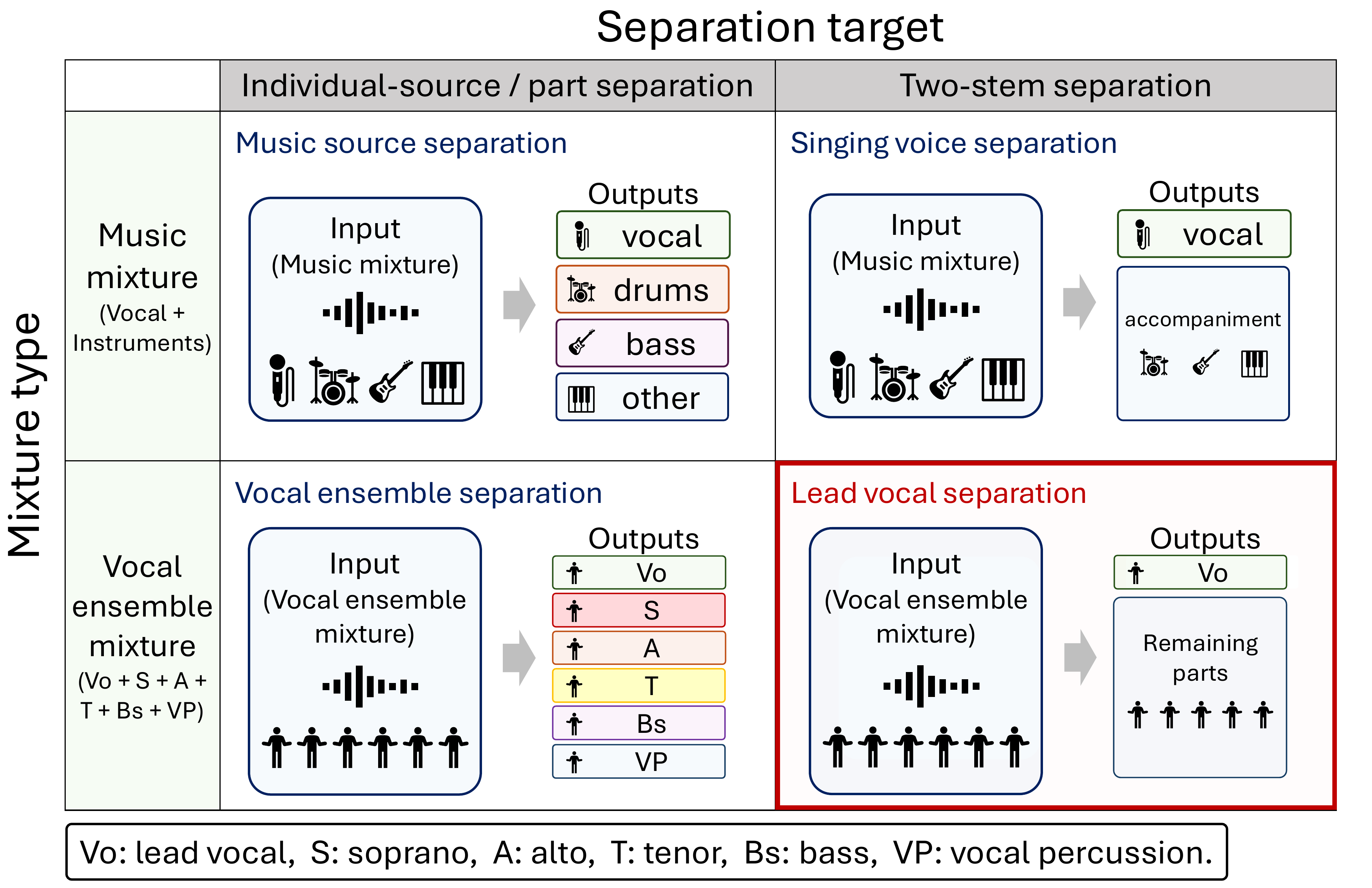}
  \caption{Relationship between Vo separation and related source separation tasks.}
  \label{fig:task-explanation}
\end{figure}

\Cref{fig:task-explanation} shows the relationship between Vo separation and related source separation tasks.
Vo separation shares the mixture type with vocal ensemble separation because both operate on vocal ensemble mixtures, while it is similar to singing voice separation in separating a melody-carrying vocal part from the remaining sources.
We briefly review vocal ensemble separation and lyric-derived auxiliary information for singing voice separation.

\subsection{Vocal Ensemble Separation} \label{sec:ves}
Vocal ensemble separation is the task of separating a vocal ensemble mixture into signals of individual vocal parts.
Research on this task has progressed largely by adapting deep neural network (DNN) architectures that have shown strong performance in other source separation tasks~\cite{Sarkar2021Interspeech,TNakamura202306ICASSP,Pan2025AI4Music,Luca2026ICASSP}.
Sarkar \textit{et al.} applied the dual-path Transformer network (DPTNet)~\cite{Chen2020Interspeech}, originally proposed for speech separation, to four-part vocal ensemble separation of soprano (S), alto (A), tenor (T), and bass (Bs)~\cite{Sarkar2021Interspeech}.
Nakamura \textit{et al.} addressed six-part vocal ensemble separation including Vo and vocal percussion (VP) by adopting DPTNet, multiresolution deep layered analysis (MRDLA)~\cite{Nakamura2021IEEEACMTASLP}, and X-UMX~\cite{Sawata2021ICASSP}.
MRDLA was originally proposed for music source separation, and X-UMX was the baseline model of the Music Demixing Challenge 2021 \cite{Mitsufuji2022FSP}, a competition on music source separation.
Hybrid Transformer Demucs~\cite{Rouard2023ICASSP}, one of the state-of-the-art music source separation models, has also been applied to VP extraction from vocal ensemble mixtures~\cite{Pan2025AI4Music}.
Separation-and-reconstruction Transformer (SepReFormer)~\cite{Shin2024NeurIPS} was originally presented for speech separation and has been adapted to six-part vocal ensemble separation~\cite{Luca2026ICASSP}. 
These studies have advanced vocal ensemble separation mainly in a mixture-only setting, where individual vocal parts are estimated from the input mixture without auxiliary information.

As an approach to vocal ensemble separation using auxiliary information, a method has been proposed that conditions the separation model on the fundamental frequency ($\Fo$) trajectories of the individual vocal parts~\cite{Petermann2020ISMIR}.
This approach shows that part-specific auxiliary information can be incorporated into vocal ensemble separation.
However, it relies on $\Fo$ trajectories that are already associated with the corresponding vocal parts.
Obtaining such part-associated $\Fo$ trajectories from a mixture is difficult because it requires not only multi-$\Fo$ estimation for overlapping voices but also assignment of the estimated trajectories to individual vocal parts over time.
Thus, using this approach as a comparison would require solving an additional part-tracking problem or using oracle $\Fo$ trajectories, making it less directly comparable to the setting considered in this paper.

\subsection{Auxiliary Information for Singing Voice Separation}
Several types of auxiliary information have been explored for singing voice separation~\cite{Gupta2022IEEEACMTASLP}, including performance videos~\cite{Montesinos2021BMVC} and lyric-derived information~\cite{Nguyen2024EUSIPCO,MeseguerBrocal2020ISMIR,SchulzeForster2021IEEEACMTASLP,Jeon2020ISMIR}.
Since the Vo part in contemporary a cappella singing typically carries the main lyrics, lyric-derived information can provide a direct cue for Vo separation.
We thus focus on lyric-derived information as auxiliary information for Vo separation.

Lyric-derived information has been used in singing voice separation under different assumptions about temporal alignment.
Jeon \textit{et al.} used time-aligned lyrics by encoding them and providing the resulting features to a separation network~\cite{Jeon2020ISMIR}.
Meseguer-Brocal and Peeters used aligned phoneme information to condition intermediate representations using FiLM~\cite{MeseguerBrocal2020ISMIR}.
Recent studies have also addressed settings where time alignment is not given in advance by estimating or jointly modeling lyric-to-audio alignment and separation~\cite{SchulzeForster2021IEEEACMTASLP,Nguyen2024EUSIPCO}.
For this first study on Vo separation from six-part vocal ensemble recordings, we use phoneme alignment as a controlled, time-synchronized form of lyric-derived information in this paper.
This allows us to evaluate the potential benefit of phoneme conditioning independently of errors introduced by automatic alignment.

\begin{figure*}[t!]
  \centering
  \includegraphics[width=\linewidth]{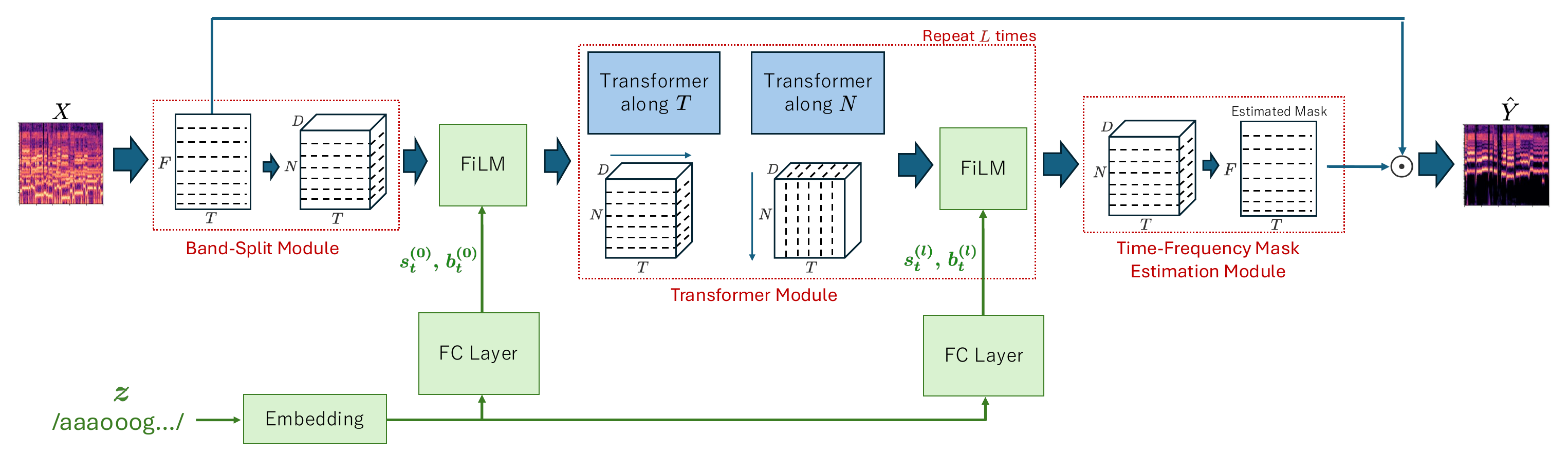}
  \caption{Network architectures of the proposed model. The green modules denote the conditioning mechanism for phoneme alignment; all other modules are the same as in the original BS-RoFormer.}
  \label{fig:models}
\end{figure*}

\section{Proposed Method}
\label{sec:proposed}

In this section, we describe the proposed Vo separation method, which incorporates phoneme alignment of the Vo part into BS-RoFormer.
We first review the network architecture of BS-RoFormer and then describe the FiLM-based conditioning mechanism for phoneme alignment.

\subsection{Network Architecture of BS-RoFormer}
\label{sec:bs-roformer}

As discussed in \Cref{sec:ves}, prior vocal ensemble separation studies have often adapted DNNs from speech or music source separation.
Following this approach, we adopted BS-RoFormer~\cite{Lu2024ICASSP}, a state-of-the-art DNN for music source separation, as the backbone network architecture for Vo separation.
In the following, we describe the network architecture adapted for Vo separation.

BS-RoFormer performs separation in the complex spectrogram domain.
It consists of a band-split module~\cite{Luo2023IEEEACMTASLP}, Transformer modules~\cite{Vaswani2017NeurIPS}, and a time-frequency mask estimation module.
Let $\bm{X}\in\C^{T\times F}$ denote the complex spectrogram obtained by applying the short-time Fourier transform (STFT) to a monaural input signal, where $T$ and $F$ are the numbers of time frames and frequency bins, respectively.
The band-split module $\mathcal{B}$ splits $\bm{X}$ into $N$ band-wise components according to predefined frequency bands and maps each component to a $D$-dimensional intermediate representation $\bm{x}_{t,n}^{(0)}\in\R^{D}$:
\begin{equation}
    \{\bm{x}_{t,n}^{(0)}\}_{t=0,n=0}^{T-1,N-1} = \mathcal{B}(\bm{X}),
\end{equation}
where $n$ denotes the frequency-band index and $N$ is the number of frequency bands.

The Transformer modules alternately apply temporal and frequency-band Transformer blocks.
The output of the $l$-th Transformer module $\mathcal{T}^{(l)}$ is given as
\begin{equation}
\{\bm{x}_{t,n}^{(l)}\}_{t=0,n=0}^{T-1,N-1}
=
\mathcal{T}^{(l)}
\left(
\{\bm{x}_{t,n}^{(l-1)}\}_{t=0,n=0}^{T-1,N-1}
\right),
\end{equation}
where $l=1,\ldots,L$ and $L$ is the number of Transformer modules.
Rotary positional embeddings (RoPE)~\cite{Su2024Neurocomputing} are used in these Transformer modules to encode relative positional information.

After $L$ Transformer modules, the mask estimation module $\mathcal{M}$ estimates a complex mask $\bm{M}\in\C^{T\times F}$ for the Vo part:
\begin{equation}
    \bm{M}
    =
    \mathcal{M}
    \left(
    \{\bm{x}_{t,n}^{(L)}\}_{t=0,n=0}^{T-1,N-1}
    \right).
\end{equation}
The separated spectrogram is then obtained as
\begin{equation}
    \hat{\bm{Y}} = \bm{M}\odot\bm{X},
\end{equation}
where $\odot$ denotes element-wise multiplication.
The estimated complex spectrogram of the remaining vocal parts is computed as $\bm{X}-\hat{\bm{Y}}$.

\subsection{Incorporating Phoneme Alignment Using FiLM}
We incorporate the phoneme alignment of the Vo part into BS-RoFormer using FiLM.
\Cref{fig:models} illustrates the architecture of the proposed network.
The proposed network keeps the input-output format of BS-RoFormer unchanged and inserts FiLM layers after the band-split module and after each Transformer module.
Each FiLM layer is conditioned on the frame-level phoneme label of the Vo part and modulates the intermediate representation at the corresponding time frame.

The phoneme alignment is represented as a frame-level sequence of phoneme-label indices $\bm{z}=[z_1,\ldots,z_T]^\top$, where $z_t\in\{0,1,\ldots,P-1\}$ denotes the phoneme-label index assigned to time frame $t$ and $P$ is the number of phoneme labels including phonemes, silence, and padding.
The silence label is assigned to valid frames in which the Vo part is not singing, whereas the padding label is used only for dummy frames introduced during batching.

Phoneme-label index $z_t$ is converted into an $E$-dimensional embedding vector:
\begin{equation}
\bm{u}_t = \mathcal{E}(z_t),
\end{equation}
where $\mathcal{E}:\{0,1,\ldots,P-1\}\rightarrow\mathbb{R}^{E}$ denotes a learnable lookup table.
We index the variables of the FiLM layers by $l=0,\ldots,L$, where $l=0$ denotes the FiLM layer after the band-split module and $l\geq 1$ denotes the FiLM layer after the $l$-th Transformer module.
For the $l$-th FiLM layer, the scale and bias vectors, $\bm{s}^{(l)}_t\in\R^{D}$ and $\bm{b}^{(l)}_t\in\R^{D}$, are generated as
\begin{equation}
\begin{bmatrix}
    \bm{s}^{(l)}_t \\
    \bm{b}^{(l)}_t
\end{bmatrix}
=
\text{FC}^{(l)}(\bm{u}_t),
\label{eq:film_gen}
\end{equation}
where $\text{FC}^{(l)}$ is a fully connected (FC) layer.
Since the phoneme label $z_t$ is defined at the frame level, the FiLM parameters $\bm{s}^{(l)}_t$ and $\bm{b}^{(l)}_t$ depend on $t$ but not on the frequency-band index $n$.
The intermediate representation is then modulated as
\begin{equation}
\tilde{\bm{x}}^{(l)}_{t,n}
=
\bm{s}^{(l)}_t \odot \bm{x}^{(l)}_{t,n} + \bm{b}^{(l)}_t.
\label{eq:film}
\end{equation}
The modulated representation $\{\tilde{\bm{x}}^{(l)}_{t,n}\}_{t=0,n=0}^{T-1,N-1}$ is passed to the subsequent module.
Since the FiLM layers modify only the intermediate representations, the conditioned model preserves the output format of the BS-RoFormer baseline and can be trained with the same loss formulation.

\section{Experiments}
\label{sec:exp}
We conducted experiments to investigate the impact of phoneme alignment on Vo separation.
We also evaluated an activity-only condition derived from the phoneme alignments to examine the effect of Vo singing/silence information.

\subsection{Data} \label{sec:exp_data}
The training and evaluation data were created using the jaCappella corpus~\cite{TNakamura202306ICASSP}.
The jaCappella corpus consists of 50 vocal ensemble pieces with six vocal parts: Vo, S, A, T, Bs, and VP.
It includes mixtures of all parts, monaural isolated recordings for each part, and musical scores.
The pieces were created by arranging Japanese children's songs and school songs whose copyright protection periods had expired, and are divided into ten subsets, each containing five pieces from a different genre.
All pieces in each subset are performed by the same set of singers for the six vocal parts.
The sampling frequency is \SI{48}{\kilo\hertz}.

For each piece, the six vocal parts were reorganized into two sources: \emph{Vo} and the remaining vocal parts, denoted as \emph{Other}.
The Vo source was the original Vo recording, and the Other source was created by mixing the S, A, T, Bs, and VP sources with equal gain.
The pieces were split into 40 training pieces and 10 evaluation pieces, with the evaluation set containing one piece from each of the ten subsets.
We also ensured that no phoneme label appeared only in the evaluation set.

The jaCappella corpus does not include phoneme alignments.
Therefore, we annotated phoneme alignment for the Vo recordings.
Initial alignments were obtained by applying dynamic time warping between phoneme sequences derived from the lyrics and the isolated Vo recordings.
Annotators with phonetics and linguistics knowledge manually corrected the initial alignments using the speech analysis software Praat\footnote{\url{https://praat.org/}}.
Our experiments used these manually corrected alignments to evaluate the potential of phoneme-alignment conditioning under the idealized assumption of accurate phoneme alignment.
During training, these annotations were converted into frame-level phoneme-label sequences.
The number of labels was $P=36$.

For the activity-only condition, the phoneme-label sequences were converted into three-label activity sequences consisting of padding, silence, and singing.
Frames labeled as padding or silence were kept unchanged, whereas frames labeled with phonemes were assigned to the singing label.

\begin{table*}[t!]
    \centering
   \caption{SI-SDRi~(\si{\decibel}) for Vo and Other obtained by Baseline, Activity-only, and Proposed}
   \footnotesize
  \begin{tabular}{cc|ccc|ccc}
        \toprule
        \multirow{2}{*}{Song name} & \multirow{2}{*}{Duration (s)} & \multicolumn{3}{c|}{Vo} & \multicolumn{3}{c}{Other} \\
        & & Baseline & Activity-only & Proposed & Baseline & Activity-only & Proposed \\
        \midrule
        Akaiboushishiroiboushi   & 66.0 & 10.47 & 15.54 & \textbf{18.29} & 7.65 & 12.61 & \textbf{15.31} \\
        Akaikutsu           & 43.8 & 9.72 & 12.08 & \textbf{13.36} & 9.11 & 11.32 & \textbf{12.46} \\
        Akatonbo          & 84.5 & 9.68 & 11.64  & \textbf{12.92} & 7.21 & 9.00 & \textbf{10.21} \\
        Anomachikonomachi      & 57.3 & 8.60 & 15.75 & \textbf{18.48} & 1.60 & 7.80 & \textbf{10.11} \\
        Aogebatoutoshi        & 75.8 & 13.71 & 17.42 & \textbf{18.14} & 11.47 & 14.75 & \textbf{15.08} \\
        Dongurikorokoro   & 75.1 & 12.95 & \textbf{16.65} & 14.52 & 9.63 & \textbf{13.11} & 11.09 \\
        Doubutsuen            & 68.9 & 9.66 & 12.72 & \textbf{17.22} & 8.16 & 11.20 & \textbf{15.53} \\
        Harugakita          & 38.2 & 3.63 & 3.92 & \textbf{9.53} & 1.64 & 2.31 & \textbf{7.33} \\
        Kisha             & 45.1 & 6.23 & \textbf{9.29} & 8.39 & 5.48 & \textbf{8.20} & 7.60 \\
        Koganemushi           & 66.8 & 14.02 & 15.45 & \textbf{17.63} & 9.51 & 10.82 & \textbf{12.91} \\
        \midrule
        Average       & - & 9.87 & 13.05 & \textbf{14.85} & 7.15 & 10.11 & \textbf{11.76}\\
        \bottomrule
  \end{tabular}
  \label{tab:results}
\end{table*}

\subsection{Compared Models}

We compared the following three models.
All models were implemented based on a public repository of music source separation\footnote{\url{https://github.com/ZFTurbo/Music-Source-Separation-Training}}.

\noindent\textbf{Baseline}: 
This model applies BS-RoFormer to Vo separation using only the mixture audio as input.

\noindent\textbf{Activity-only}: This model uses the Vo activity as auxiliary information. It was used to examine whether Vo activity information alone accounts for the effect of phoneme alignment.
The number of conditioning labels was $P=3$, and the embedding dimension was set to $E=36$.

\noindent\textbf{Proposed}: This model is the proposed model described in \Cref{sec:proposed}. It uses the frame-level phoneme alignment of the Vo part as auxiliary information.
The number of conditioning labels and the embedding dimension were set to $P=36$ and $E=36$, respectively.

For all models, the BS-RoFormer backbone was based on the public BS-RoFormer configuration for  MUSDB18\footnote{\url{https://github.com/ZFTurbo/Music-Source-Separation-Training/blob/main/configs/config_musdb18_bs_roformer.yaml}}.
In the band-split module, the number of frequency bands was set to $N=62$.
In the Transformer modules, we set $D=192$ and $L=6$, and the number of attention heads was set to 8.
The STFT window length was 2048 samples (\SI{42.7}{\milli\second}), the frame shift was 512 samples (\SI{10.7}{\milli\second}), and a Hann window was used.
All other backbone hyperparameters were set according to the same configuration.

\subsection{Training Conditions}

All models were trained under the same conditions.
The batch size was set to 8.
Each sample was created by randomly cropping a segment of \num{131584} samples (\SI{2.74}{\second}) from the Vo and Other source signals during minibatch generation.
As data augmentation, the gains of Vo and Other were independently and randomly changed within the range from 0.5 to 1.5.

As the loss function, we used the mean absolute error between the estimated and reference Vo signals, together with the multi-resolution STFT loss \cite{Guso2022ICASSP}.
The multi-resolution STFT loss was computed as the sum of the mean absolute errors between the complex spectrograms of the estimated and reference signals obtained using Hann windows with window lengths of $256,512,1024,2048,$ and $4096$ samples ($5.3,10.7,21.3,42.7,$ and $85.3$~\si{\milli\second}).
The frame shift was set to 147 samples (\SI{3.1}{\milli\second}) for all window lengths.
Optimization was performed using Adam with a learning rate of $\num{5e-5}$ and momentum decay rates of $\beta_1=0.9$ and $\beta_2=0.999$.
All models were trained for \num{1000} epochs with \num{1000} minibatch updates per epoch.
The models obtained at the last epoch were used for evaluation.

\subsection{Results}

\Cref{tab:results} shows the separation performance of \textbf{Baseline}, \textbf{Activity-only}, and \textbf{Proposed}.
Separation performance was evaluated using the scale-invariant source-to-distortion ratio improvement (SI-SDRi), which measures the improvement over the input mixture~\cite{LeRoux2019ICASSP}.

\textbf{Proposed} outperformed \textbf{Baseline} for all evaluation pieces in both Vo and Other.
It also achieved the highest average SI-SDRi for both outputs.
These results demonstrate that phoneme alignment improves both Vo extraction and Vo removal, yielding cleaner estimates of both parts.

\textbf{Activity-only} achieved the second-best average SI-SDRi for both Vo and Other, while outperforming \textbf{Baseline} for all evaluation pieces.
Together with the higher average performance of \textbf{Proposed}, this result shows that the benefit of phoneme alignment is not merely due to identifying when the Vo part is active.
One possible reason is that phoneme labels group Vo-active frames into phonetic categories, so that frames with the same phoneme label are treated as belonging to the same class rather than simply as singing frames.
Such categorical relations between frames should provide additional information for separating the Vo part from the remaining vocal parts.

While \textbf{Proposed} achieved the highest average SI-SDRi, \textbf{Activity-only} gave higher SI-SDRi than \textbf{Proposed} for Dongurikorokoro and Kisha, which were performed by the same singer group.
This observation suggests that singer- or ensemble-configuration-dependent factors may affect the benefit of phoneme-label conditioning.
A more detailed analysis of this effect is left for future work.

\subsection{Phoneme-Overlap Analysis}
To examine how phoneme alignment contributes to Vo separation, we analyzed separation performance under different phoneme-overlap conditions between Vo and the other parts.
For each Vo-active frame, we counted the number of other parts whose phoneme label matched that of Vo, and grouped frames according to this count.

\Cref{tab:overlap} shows the duration-weighted average SI-SDRi for each phoneme-overlap condition. The duration-weighted average is calculated by computing SI-SDRi for each song using only the samples that satisfy each phoneme-overlap condition, and then averaging the song-wise SI-SDRi values with weights proportional to the duration of the selected samples.

\textbf{Proposed} achieved higher SI-SDRi than \textbf{Baseline} across all conditions, and the improvement was relatively consistent across conditions.
This shows that the effect of phoneme alignment is not limited to frames where Vo and the other parts sing different phonemes.
In contrast, the gap between \textbf{Proposed} and \textbf{Activity-only} tended to decrease as the number of overlapping vocal parts increased.
This trend is consistent with the intuition that phoneme-category information is most useful when it helps distinguish the Vo part from the other parts at the phoneme level, whereas its advantage over Vo activity becomes smaller when more vocal parts share the same phoneme category.

\begin{table}[t]
    \centering
   \caption{Duration-weighted average SI-SDRi~(\si{\decibel}) for Vo separation under different phoneme-overlap conditions.
            The proportion represents the duration of each condition divided by the total duration}
   \footnotesize
  \begin{tabular}{cc|ccc}
        \toprule
        \# of overlaps & Proportion~(\%) & Baseline & Activity-only & Proposed \\
        \midrule
         0  & 40.41 & 12.26 & 14.83 & \textbf{18.32} \\
         1  & 14.68 & 11.85 & 15.01 & \textbf{17.21} \\
         2  & 8.67 & 9.00 & 11.20 & \textbf{13.16} \\
         3  & 19.75 & 9.25 & 11.74 & \textbf{12.15} \\
         4  & 16.48 & 9.19 & 12.23 & \textbf{13.83} \\
        \bottomrule
  \end{tabular}
  \label{tab:overlap}
\end{table}

\section{Conclusion}

In this paper, we proposed a Vo separation method using phoneme alignment as auxiliary information.
The proposed method incorporates frame-level phoneme labels of the Vo part into BS-RoFormer through FiLM-based conditioning.
Experimental results on the jaCappella corpus showed that the proposed method improved SI-SDRi over the audio-only baseline for both Vo and Other, demonstrating that phoneme alignment is effective for both Vo extraction and Vo removal.
The activity-only conditioning also improved separation performance, but the proposed method achieved higher average SI-SDRi.
This result suggests that the benefit of phoneme alignment is not merely due to Vo singing/silence activity, but also to finer-grained phoneme-label information.
The phoneme-overlap analysis further showed that the improvement over the baseline was observed across different overlap conditions, while the advantage over activity-only became smaller when more vocal parts shared the same phoneme category.
Future work will investigate methods that use lyrics without accurately aligned phoneme labels to improve practical usability.
Subjective listening evaluation is also left for future work.

\section*{Acknowledgment}
This work was supported by JSPS KAKENHI Grant Number JP23K28108.

\printbibliography

@inproceedings{Chen2020Interspeech,
  author = {Chen, J. and Mao, Q. and Liu, D.},
  booktitle = {Proc. Interspeech},
  month = {October},
  pages = {2642--2646},
  title = {Dual-Path {Transformer} Network: {Direct} Context-Aware Modeling for End-to-End Monaural Speech Separation},
  year = {2020}
}

@article{Nakamura2021IEEEACMTASLP,
  author = {Nakamura, T. and Kozuka, S. and Saruwatari, H.},
  journal = {IEEE/ACM Trans. Audio, Speech, Lang. Process.},
  pages = {1687--1701},
  title = {Time-Domain Audio Source Separation With Neural Networks Based on Multiresolution Analysis},
  volume = {29},
  year = {2021}
}

@article{Mitsufuji2022FSP,
  author = {Mitsufuji, Y. and Fabbro, G. and Uhlich, S. and St{\"o}ter, {F.-R.} and D{\'e}fossez, A. and Kim, M. and Choi, W. and Yu, C.-Y. and Cheuk, K.-W.},
  journal = {Front. Signal Process.},
  title = {Music Demixing Challenge 2021},
  volume = {1},
  year = {2022}
}

@inproceedings{Montesinos2021BMVC,
  author = {Montesinos, J.~F. and Kadandale, V.~S. and Haro, G.},
  booktitle = {Proc. Brit. Mach. Vis. Conf.},
  title = {{A cappella: A}udio-visual Singing Voice Separation},
  year = {2021}
}

@inproceedings{Sarkar2021Interspeech,
  author = {Sarkar, S. and Benetos, E. and Sandler, M.},
  booktitle = {Proc. Interspeech},
  pages = {3515--3519},
  title = {Vocal Harmony Separation Using Time-Domain Neural Networks},
  year = {2021}
}

@inproceedings{LeRoux2019ICASSP,
  author = {{Le Roux}, J. and Wisdom, S. and Erdogan, H. and Hershey, J.~R.},
  booktitle = {Proc. IEEE Int. Conf. Acoust., Speech, Signal Process.},
  pages = {626--630},
  title = {{SDR} -- half-baked or well-done?},
  year = {2019}
}

@inproceedings{Petermann2020ISMIR,
  author = {Petermann, D. and Chandna, P. and Cuesta, H. and Bonada, J. and G{\'o}mez, E.},
  booktitle = {Proc. Int. Soc. Music Inf. Retr. Conf.},
  pages = {733--739},
  title = {Deep Learning Based Source Separation Applied to Choir Ensembles},
  year = {2020}
}

@inproceedings{TNakamura202306ICASSP,
  author = {Nakamura, Tomohiko and Takamichi, Shinnosuke and Tanji, Naoko and Fukayama, Satoru and Saruwatari, Hiroshi},
  booktitle = {Proc. IEEE Int. Conf. Acoust., Speech, Signal Process.},
  lang = {en},
  month = {June},
  title = {{jaCappella} Corpus: {A} {Japanese} A Cappella Vocal Ensemble Corpus},
  year = {2023}
}

@article{Luo2023IEEEACMTASLP,
  author = {Luo, Yi and Yu, Jianwei},
  journal = {IEEE/ACM Trans. Audio, Speech, Lang. Process.},
  pages = {1893--1901},
  title = {Music Source Separation With Band-Split {RNN}},
  volume = {31},
  year = {2023}
}

@inproceedings{Lu2024ICASSP,
  author = {Lu, Wei-Tsung and Wang, Ju-Chiang and Kong, Qiuqiang and Hung, Yun-Ning},
  booktitle = {Proc. IEEE Int. Conf. Acoust., Speech, Signal Process.},
  pages = {481-485},
  title = {Music Source Separation With Band-Split {RoPE} {Transformer}},
  year = {2024}
}

@inproceedings{Vaswani2017NeurIPS,
  author = {Vaswani, Ashish and Shazeer, Noam and Parmar, Niki and Uszkoreit, Jakob and Jones, Llion and Gomez, Aidan N. and Kaiser, \L{}ukasz and Polosukhin, Illia},
  booktitle = {Proc. Adv. Neural Inf. Process. Syst.},
  pages = {6000--6010},
  title = {Attention is all you need},
  year = {2017}
}

@article{Su2024Neurocomputing,
  author = {Jianlin Su and Murtadha Ahmed and Yu Lu and Shengfeng Pan and Bo Wen and Yunfeng Liu},
  journal = {Neurocomputing},
  number = {127063},
  title = {{RoFormer:} Enhanced transformer with Rotary Position Embedding},
  volume = {568},
  year = {2024}
}

@inproceedings{Sawata2021ICASSP,
  author = {Sawata, Ryosuke and Uhlich, Stefan and Takahashi, Shusuke and Mitsufuji, Yuki},
  booktitle = {Proc. IEEE Int. Conf. Acoust., Speech, Signal Process.},
  pages = {51--55},
  title = {All For One And One For All: {Improving} Music Separation By Bridging Networks},
  year = {2021}
}

@inproceedings{Perez2018AAAI,
  author = {Perez, Ethan and Strub, Florian and de Vries, Harm and Dumoulin, Vincent and Courville, Aaron},
  booktitle = {Proc. AAAI Conf. Artif. Intell.},
  number = {483},
  title = {{FiLM}: {Visual} reasoning with a general conditioning layer},
  year = {2018}
}

@inproceedings{MeseguerBrocal2020ISMIR,
  author = {Meseguer-Brocal, Gabriel and Peeters, Geoffroy},
  booktitle = {Proc. Int. Soc. Music Inf. Retr. Conf.},
  title = {Content Based Singing Voice Source Separation via Strong Conditioning Using Aligned Phonemes},
  year = {2020}
}

@inproceedings{Nguyen2024EUSIPCO,
  author = {Nguyen, Th\'{e}o and Teytaut, Yann and Roebel, Axel},
  booktitle = {Proc. Eur. Signal Process. Conf.},
  pages = {366--370},
  title = {On Strategies to Exploit Dependencies Between Singing Voice Alignment and Separation},
  year = {2024}
}

@article{SchulzeForster2021IEEEACMTASLP,
  author = {Schulze-Forster, Kilian and Doire, Clement S. J. and Richard, Ga\"{e}l and Badeau, Roland},
  journal = {IEEE/ACM Trans. Audio, Speech, Lang. Process.},
  pages = {2382--2395},
  title = {Phoneme Level Lyrics Alignment and Text-Informed Singing Voice Separation},
  volume = {29},
  year = {2021}
}

@inproceedings{Jeon2020ISMIR,
  author = {Jeon, Chang-Bin and Choi, Hyeong-Seok and Lee, Kyogu},
  booktitle = {Proc. Int. Soc. Music Inf. Retr. Conf.},
  pages = {685--692},
  title = {Exploring Aligned Lyrics-Informed Singing Voice Separation},
  year = {2020}
}

@inproceedings{Pan2025AI4Music,
  author = {Ting-Yu Pan and
Kexin Phyllis Ju and
Hao-Wen Dong},
  booktitle = {NeurIPS Workshop on Artificial Intelligence for Music},
  title = {{ACappellaSet:} {A} Multilingual A Cappella Dataset for Source Separation and {AI}-assisted Rehearsal Tools},
  year = {2025}
}

@inproceedings{Luca2026ICASSP,
  author = {Lanzendorfer, Luca A. and Pinkl, Constantin and Gr\"{o}tschla, Florian},
  booktitle = {Proc. IEEE Int. Conf. Acoust., Speech, Signal Process.},
  pages = {14597--14601},
  title = {Source Separation For A Cappella Music},
  year = {2026}
}

@article{Gupta2022IEEEACMTASLP,
  author = {Gupta, Chitralekha and Li, Haizhou and Goto, Masataka},
  journal = {IEEE/ACM Trans. Audio, Speech, Lang. Process.},
  pages = {2422--2451},
  title = {Deep Learning Approaches in Topics of Singing Information Processing},
  volume = {30},
  year = {2022}
}

@inproceedings{Rouard2023ICASSP,
  author = {Rouard, Simon and Massa, Francisco and Défossez, Alexandre},
  booktitle = {Proc. IEEE Int. Conf. Acoust., Speech, Signal Process.},
  title = {Hybrid {Transformers} for Music Source Separation},
  year = {2023}
}

@inproceedings{Shin2024NeurIPS,
  author = {Shin, Ui-Hyeop and Lee, Sangyoun and Kim, Taehan and Park, Hyung-Min},
  booktitle = {Proc. Adv. Neural Inf. Process. Syst.},
  pages = {52215--52240},
  title = {Separate and Reconstruct: {Asymmetric} Encoder-Decoder for Speech Separation},
  volume = {37},
  year = {2024}
}

@inproceedings{Guso2022ICASSP,
  author = {Gus\'{o}, Enric and Pons, Jordi and Pascual, Santiago and Serr\'{a}, Joan},
  booktitle = {Proc. IEEE Int. Conf. Acoust., Speech, Signal Process.},
  pages = {306--310},
  title = {On Loss Functions and Evaluation Metrics for Music Source Separation},
  year = {2022}
}

@article{Duchan2007Collegiate,
  author = {Duchan, Joshua S.},
  journal = {American Music},
  number = {4},
  pages = {477--506},
  title = {Collegiate a Cappella: {Emulation} and Originality},
  volume = {25},
  year = {2007}
}

@book{acappella101,
  author = {Dietz, Rob},
  publisher = {Hal Leonard},
  title = {{A} Cappella 101: {A} Beginner's Guide to Contemporary A Cappella Singing},
  year = {2022}
}

\end{document}